\documentclass[12pt]{amsart}
\let\useblackboard=\iftrue
\usepackage{amssymb}
\usepackage{amsfonts}
\usepackage{amsthm}
\numberwithin{equation}{section}
\theoremstyle{plain}
\newtheorem{theorem}{Theorem}[section]

\newtheorem{claim}[theorem]{Claim}

\theoremstyle{definition}

\useblackboard
\font\blackboard=msbm10 scaled \magstep1
\font\blackboards=msbm7
\font\blackboardss=msbm5
\newfam\black
\textfont\black=\blackboard
\scriptfont\black=\blackboards
\scriptscriptfont\black=\blackboardss
\def\Bbb#1{{\fam\black\relax#1}}
\else
\def\Bbb{\bf}
\fi
\def\yboxit#1#2{\vbox{\hrule height #1 \hbox{\vrule width #1
\vbox{#2}\vrule width #1 }\hrule height #1 }}
\def\fillbox#1{\hbox to #1{\vbox to #1{\vfil}\hfil}}
\def\ybox{{\lower 1.3pt \yboxit{0.4pt}{\fillbox{8pt}}\hskip-0.2pt}}
\def\mapr{\mathop{\longrightarrow}\limits}

\def\grade{\varphi}
\def\cal{\mathcal}
\def \I {I}
\def \II {II}

\def \Tr {{\rm Tr}}

\def \ap {\alpha'}
\def\BC{\Bbb{C}}
\def\BP{\Bbb{P}}
\def\P{\Bbb{P}}

\def\BZ{\Bbb{Z}}
\def\p{\partial}

\def\rk{{\rm rk\ }}

\def\CN{{\cal N}}
\def\CO{{\cal O}}

\begin{document}

\title{D-Branes and $\CN=1$ Supersymmetry}

\author[M.R. Douglas]{Michael R. Douglas}
\address[Michael R. Douglas]{
Dept. of Physics and Astronomy, Rutgers University,
Piscataway, NJ 08855 USA {\it and}
I.H.E.S., Le Bois-Marie, Bures-sur-Yvette 91440 France}
\email{mrd@physics.rutgers.edu}
\begin{abstract}
We discuss the recent proposal that BPS D-branes in Calabi-Yau
compactification of type \II\ string theory are $\Pi$-stable
objects in the derived category of coherent sheaves.

To appear in the proceedings of Strings 2001, Mumbai, India.
\end{abstract}

\date{April 30, 2001}
\maketitle
\section{Introduction}\label{S:intro}
Superstring theory has made enormous progress over the last decade.
We now think of it as part of a larger framework, string/M theory,
which has given us some nonperturbative understanding and ability to
address questions of relatively direct physical interest, such as the
qualiative dynamics of strongly coupled gauge theory and the origin
of black hole entropy.

Less progress has been made on the primary question from the 80's,
namely to make contact with particle physics phenomenology.
Nevertheless, I believe this will be the central problem for the
theory over the coming decade, partly because our improved
understanding does bear on the difficulties encountered then, and even
more (one hopes) because of the prospect of entirely new physics to be
discovered at LHC, Fermilab and elsewhere.  Many physicists believe that
supersymmetry will be found at the energies to be probed there, in which
case the central problem will be to understand backgrounds of
string/M theory with $\CN=1$ supersymmetry at low energies.

Any systematic approach to these questions requires techniques in
which all, or at least some large class of backgrounds, can be
studied.  One might be lucky and find that a particularly simple
background turns out to show interesting parallels with observed
physics, but without the larger picture, one will not know how much
significance to ascribe to this.  Furthermore, our experience with
duality suggests that it can often be easier to understand and get
exact results for all backgrounds of a theory than for any given
background; the Seiberg-Witten solution of $\CN=2$ Yang-Mills theory
provides a compelling example.

Looking back over the results from string/M theory duality, it seems
fair to say that a good understanding was achieved of four
dimensional backgrounds with $\CN=2$ supersymmetry, and much less
was achieved for $\CN=1$ supersymmetry.  Although we are far from a
complete classification of backgrounds with $\CN=2$ supersymmetry,
the basic test of this statement is that when one considers
a particular $\CN=2$ construction in enough depth, one is always able
to find dualities or connections to other constructions and
link it into a larger picture.

The broadest and most useful standard construction of $\CN=2$
backgrounds appears to be the compactification of type \II\
superstring theory on three (complex) dimensional Calabi-Yau
manifolds.  This provides a geometric picture and one can take
advantage of the large body of work constructing and classifying these
manifolds.  Furthermore, by using mirror symmetry, one can compute the
basic observable of $\CN=2$ supergravity, the prepotential governing
the dynamics of vector multiplets and the central charges of BPS
states, including all world-sheet instanton corrections.  This
provides exact results in an {\it a priori} highly stringy and
nongeometric regime.  Further dualities such as type \II-heterotic
allow reinterpreting these instanton corrections as space-time
instanton corrections \cite{KachruVafa}, and all of the interesting
physics of $\CN=2$ compactification discovered so far can be realized
in this framework.

Because the structure of $\CN=2$ supergravity fits so well with the
geometry of these manifolds \cite{Strominger}, it is even tempting
to conjecture that all $\CN=2$ compactifications can be realized (in
some dual picture) as type \II\ on Calabi-Yau (perhaps with simple
generalizations such as adding discrete torsion).  Since the
Calabi-Yau manifolds themselves are not classified, it is hard to
evaluate this conjecture at present, but one might imagine finding some
general argument which given an $\CN=2$ compactification produces the
appopriate Calabi-Yau.  If this turned out to be true, we would have
a clear sense in which all $\CN=2$ compactifications were classifiable.

$\CN=1$ compactification is more difficult to study for many reasons.
These models (especially the phenomenologically interesting ones) can
break supersymmetry spontaneously; if not, the supersymmetric vacua
are generically isolated.  Thus, the strategy of matching moduli
spaces of dual pairs which works well for $\CN=2$ will not work
in all cases.

Experience with $\CN=1$ suggests however that there are many models
with moduli spaces, and that a rich picture of dualities should
already exist for these.  In a larger picture, we might find
additional relations between this subset of theories and those with
isolated or no supersymmetric vacua, bringing these cases within reach.

Thus we should look for some standard construction of $\CN=1$
compactifications of string theory, which ideally would be geometric
and share the good properties of type \II\ on Calabi-Yau.  The
traditional answer to this question is to compactify the heterotic
string on a Calabi-Yau threefold.  This depends on the additional
choice of a vector bundle with structure group $G\subset E_8\times E_8$
or $G\subset Spin(32)/\BZ_2$ and a gauge field solving the hermitian
Yang-Mills equations \cite{GSW}.  This additional choice is geometric and
involves both topological parameters (the Chern or K theory class) and
holomorphic parameters, leading to an interesting field content,
superpotential and symmetry breaking structure.

One general problem with using this as our standard construction of
$\CN=1$ compactifications is that in general these results obtain both
world-sheet and space-time instanton corrections, and there is no
clear limit which keeps a controllable subset of the stringy
corrections.  Even the classical and large volume limit would be
interesting to have under control, but the larger problem is that no
clear picture of the set of such bundles is to be found in the physics
literature, which relies almost entirely on {\it ad hoc} constructions.
(The construction of \cite{FMW} is not {\it ad hoc} and is a definite
step forward, but it does not describe all bundles.)

One can try enlisting mathematical help for this problem, but the
specific question we just asked, what are all the bundles satisfying
certain topological constraints (the anomaly matching conditions), is
generally considered difficult.  On the other hand, the mathematicians
do know a great deal about this type of problem, it is just not in the
predigested form one might like.

Two other candidate standard constructions have emerged in more recent
times, F theory compactified on Calabi-Yau fourfolds \cite{Vafa}, and
type \I\ theory on Calabi-Yau threefolds \cite{Sagnotti}.  Both
take advantage of the equivalence between branes and gauge
field configurations which is the hallmark of the Dirichlet brane and
which was already exemplified in Witten's work on small instantons
\cite{Witten}: the moduli space of a configuration of Dirichlet
$5$-branes and $9$-branes in flat space, is equivalent to the moduli
space of instantons, and furthermore their world-volume gauge theory
contains precisely the ADHM construction of the moduli space and the
instantons themselves, making a powerful but rather abstruse
mathematical construction quite concrete and usable, as has been
demonstrated in numerous recent works, such as \cite{Dorey}.

This relation suggests that type \I\ on Calabi-Yau could serve as a
better standard construction, and that the power of the Dirichlet brane
to turn complicated mathematics into physics might allow us to proceed
much farther in this direction.

Over the last two years, this hope has been realized in the related
but somewhat simpler problem of Dirichlet branes in type \II\
compactification on Calabi-Yau threefolds.  In the classical (zero
string coupling or sphere and disk world-sheet) limit, the type \II\
and type \I\ problems are very closely related: one can obtain type
\I\ models and type \II\ orientifolds from type \II\ with branes by a
second step of quotient by a world-sheet orientation reversing symmetry.

Our work so far has concentrated on the classical type \II\ problem,
both from the desire to focus on the essentials of the problem, but
also because this liberates us from the tadpole/anomaly matching
constraints (one can consider D-branes which sit at points in
Minkowski space, which at the classical level is essentially the same
problem as space-filling branes) and allows putting the problem in a
larger context.  In particular, it turns out to be very useful to
start with a finite set of generating D-branes, which individually
would not have satisfied tadpole cancellation, and build up the
spectrum as bound states of arbitrary numbers of these.

On the other hand, we work with finite $\alpha'$, so the problem
retains highly nontrivial instanton corrections (sphere and disk
world-sheet instantons).  Thus if we could find a construction which
described all brane configurations in this limit, we would have made a
good start on the problem of finding a standard construction of
$\CN=1$ models.

In \cite{DCS}, building on \cite{BDLR,DD,Doug,DFR,DFRtwo}, we have
found a simple characterization of all the BPS branes on a Calabi-Yau,
which works in the stringy regime and predicts the dependence of the
spectrum on the CY moduli (i.e. lines of marginal stability):

\begin{claim}
A (B-type) BPS brane on the CY manifold $M$ is a $\Pi$-stable
object in the derived category of coherent sheaves on $M$.
\end{claim}

This proposal relies on some rather unfamiliar mathematics, which in
this context first appeared in Kontsevich's homological mirror
symmetry proposal \cite{Kontsevich}, and we will not be able to
explain it in detail here.  What we will try to do is explain the
basic ideas, and why this mathematics is actually very pertinent both
for the physics of branes and for the general problem of studying
$\CN=1$ supersymmetric gauge theories.  After this general
introduction, we will give an example of a Gepner model boundary state
(constructed by Recknagel and Schomerus \cite{RS}) which turns out to
be one of the ``nonclassical'' objects of the derived category.

\section{The mathematics of $\CN=1$ supersymmetry}

Consider a BPS brane in a $d=3+1$, $\CN=2$ compactification which
extends in the Minkowski dimensions.  Its world-volume theory will be
an $\CN=1$ gauge theory, which given our assumptions will have a
$\prod U(n_i)$ gauge group, chiral matter in bifundamentals, an action
which can be written as a single trace, and can be treated
classically.  Its moduli space of supersymmetric vacua will be a
moduli space of BPS brane configurations.

Study of explicit stringy constructions such as boundary states in
orbifolds and Gepner models suggests a generalization of this
idea.  Suppose we take a finite number of generating branes $B_i$ and
write the general world-volume theory with $N_i$ copies of the brane
$B_i$.  If all of these branes preserve the same $\CN=1$, everything
we just said will still be true, and we can find brane configurations
with RR charge $\sum_i N_i [B_i]$ (we use the notation $[B]$ for the
charge or K theory class of the brane $B$) by finding supersymmetric
vacua of the combined theory with unbroken $U(1)$ gauge symmetry (more
unbroken symmetry means we do not have a single bound state).  Since
it is possible to match the RR charge (or K theory class) of any brane
configuration using a finite set of generating branes, and there is no
obvious reason that this cannot realize all BPS configurations with
this charge, this suggests that we try to use this as our general
construction.  One can gain further simplicity by taking ``rigid''
branes as the $B_i$ (i.e. with no moduli space); all chiral matter
will then appear as open strings between branes, so the bound state
moduli space will be explicitly constructed by the gauge theory.

These assumptions all hold for orbifold models, but more generally
(e.g. in Gepner models), one needs to use generating branes which do
not all preserve the same $\CN=1$.  The $\CN=2$ superalgebra contains
a continuous family of $\CN=1$ subalgebras parameterized by a $U(1)$;
a BPS brane with central charge $Z$ will leave unbroken the $\CN=1$
subalgebra $Q = e^{i\pi\grade}\bar Q$ with $\pi\grade$ equal to the phase
of $Z$.  Unless the phase for each $B_i$ is the same, these
combinations will break supersymmetry.

However, at least in cases where the collection of branes actually
does decay to a BPS bound state, this might be expected to be a
spontaneous breaking of the $\CN=1$ supersymmetry of the final state.
We will pursue this assumption, eventually finding its justification
in string theory.

We now recall the structure of a general $d=4$, $\CN=1$ theory which
is relevant for the problem of finding supersymmetric vacua.  This is
the gauge group $G$, the spectrum of chiral multiplets $\phi_i$, the
superpotential $W$, and finally the D-flatness conditions, which
almost follow from the previous data but require the specification of
a real Fayet-Iliopoulos parameter for each $U(1)$ factor in $G$.  In
terms of this data, the problem of finding supersymmetric vacua splits
into a holomorphic part, finding the solutions of F-flatness $\p W/\p
\phi_i=0$ up to complex gauge equivalence, and then within each gauge
equivalence class finding a solution of the D-flatness conditions.

The first important point we want to make is that this paradigm
can serve as a general approach to
all problems of finding BPS D-branes.  Let us consider what at first
looks like the opposite limit to the one we consider, the large volume
limit, in which the brane world-volume does not have a finite number
of fields but is actually a higher dimensional gauge theory.  In this
limit, for B-type branes,
the problem is to solve the hermitian Yang-Mills equations,
\begin{eqnarray}
F_{ij} dz^i dz^j = 0 \label{eq:twozero} \\
F_{i\bar j} dz^i d\bar z^j \wedge \omega^{d-1} = c \omega^d
\label{eq:oneone}
\end{eqnarray}
where $\omega$ is the K\"ahler form and $d$ is the complex dimension
of the space.

As is well known, these are hard equations to solve explicitly, and it
is better to proceed as follows: first, find a holomorphic bundle $E$,
which can be considered as a solution of (\ref{eq:twozero}); second,
appeal to the theorems of Donaldson and Uhlenbeck-Yau \cite{DK}, which state
necessary and sufficient conditions (which we quote below) for
(\ref{eq:oneone}) to admit a solution.

This two-step procedure is precisely an infinite dimensional analog of
the two-step procedure for finding supersymmetric vacua: the equation
(\ref{eq:twozero}) is precisely $\p W/\p \phi_i=0$ where $W$ can be
taken to be the holomorphic Chern-Simons action, while
(\ref{eq:oneone}) is precisely the D-flatness condition partner to
local complex gauge transformations on the bundle $E$.

This raises the possibility that a unified procedure can cover both
the large volume limit and the approach discussed above of finding
bound states of generating branes.  A further indication that this
should be true is the ``decoupling statement'' of BDLR \cite{BDLR}.
This states that the F flatness part of the problem is independent of
half of the Calabi-Yau moduli, depending only on complex structure
moduli for B-type branes, and only on (stringy) K\"ahler moduli for
A-type branes.  In some sense this is implicit in the statement that
the superpotential is computable in topologically twisted open string
theory (e.g. see \cite{Witten-top,BCOV}), but in BDLR it was realized
that this implies that the holomorphic structure of B brane
world-volume theories, even in the stringy regime, should be exactly
computable at large volume.

\section{F flatness}

The decoupling statement predicts that there should be some very
direct equivalence between F-flat configurations of supersymmetric
gauge theories describing bound states of branes, for example the
quiver theories of \cite{DGM}, and holomorphic bundles (or coherent
sheaves \cite{HarveyMoore}) on the corresponding resolved space.  Such
an equivalence was first noticed (in the physics literature) in
\cite{DFRtwo} which studied the $\BC^3/\BZ_3$ orbifold in depth, and
it turns out to have a long mathematical history.  Indeed, the usual
$\BC^3/\BZ_3$ quiver theory precisely contains the standard
mathematical representation of the moduli space of coherent sheaves on
$\BC\BP^2$ (which is the relevant part of the resolved space), as
given by a theorem of Beilinson \cite{Beilinson} which actually
slightly predates the ADHM construction.  This theorem has already
been generalized to any $\BC^3/\Gamma$ orbifold in mathematical work
on the generalized McKay correspondence \cite{Reid} and turns out to
provide a very useful part of the story for compact Calabi-Yau
manifolds as well, at least for the cases with a Gepner model
realization, since the Gepner model can be thought of as a
Landau-Ginzburg orbifold $\BC^5/\Gamma$ and the same technology
applied \cite{DD,GJ,Mayr,Tomas}.

Although similar in spirit to constructions of sheaves which have been
used by physicists, the relations given in these theorems are better,
primarily because they are one-to-one: if a given topological class of
sheaf can be constructed, all of the sheaves in this class can be
constructed, and each sheaf corresponds to a unique configuration (up
to complex gauge equivalence) of the gauge theory.  Thus these
theorems validate the decoupling statement and provide a faithful
translation of the problem of finding bundles into gauge theory terms,
just like the original small instanton construction.

It turns out, however, that there is a deeper subtlety in the relation
between gauge theory configurations and sheaves: not all sheaves can
be constructed as bound states of branes; one needs antibranes as
well.

This becomes apparent on considering the large volume interpretation
of the fractional branes.  For $\BC^3/\BZ_3$ this was first found by
Diaconescu and Gomis \cite{DG}, and since duplicated by the many other
considerations we just described:
\begin{eqnarray}
B_1 \cong \CO_{\P^2}(-1) \\
B_2 \cong \bar \Omega_{\P^2}(1) \\
B_3 \cong \CO_{\P^2} .
\end{eqnarray}
The topological (K theory) class of a D-brane in this problem is
specified by three integers which could be regarded as D$0$, D$2$ and
D$4$-brane charge, or just as well expressed in the basis provided by
the charges of these three fractional branes $[B_i]$
(in fact all charges are
integral in the latter basis).  Thus the minimal condition for a brane
to be a bound state of fractional branes is that its charges be
nonnegative in the fractional brane basis.  It is easy to find
counterexamples, though, starting with the D$2$-brane:
\begin{equation}
[\CO_\Sigma] = [\CO_{\P^2}] - [\CO_{\P^2}(-1)].
\end{equation}
Although this directly contradicts a naive form of the decoupling statement,
the physical resolution of this point is just to allow bound states of
fractional branes and their antibranes.

There is a far more general argument that one cannot avoid talking
about bound states of branes and antibranes in these problems.  The
essential point is that the question of what is a ``brane'' and what
is an ``antibrane'' actually depends on K\"ahler moduli.  Except in
the special case of the literal antibrane $\bar B$ to a brane $B$
(i.e. its orientation reversal), the only principled way to make such
a distinction is to ask whether two objects preserve the same $\CN=1$
supersymmetry or not.  As we discussed, this will be true if their BPS
central charges have the same phase.

However, in $\CN=2$ supersymmetry, the BPS central charges depend on
the vector multiplet moduli (here the stringy K\"ahler moduli) in a
rather complicated way, determined by the $\CN=2$ prepotential.  By
looking at examples, one quickly finds that two branes with different
RR charge (K theory class) can have aligned BPS central charge at one
point in moduli space, and anti-aligned at another, continuously
interpolating along the path in between.  In particular, while in the
large volume limit the central charge is dominated by the brane of
highest dimension and thus all branes made from coherent sheaves
preserve the same $\CN=1$ supersymmetry,\footnote{ This assumes that
the volume in string units is much larger than any of the RR charges
of the branes.} elsewhere in K\"ahler moduli space this will no longer
be true.  The $\BC^3/\BZ_3$ results we just quoted are the simplest
example: at large volume $B_1$ and $B_3$ are ``branes'' while $B_2$ is
an ``antibrane,'' while at the orbifold point $\BZ_3$ symmetry
guarantees that all three have the same central charge and are
``branes.''

In large volume terms, it is not natural to restrict attention only to
bundles or to coherent sheaves; one must consider a larger class of
objects.  This is rather more apparent in the A brane picture, in
which continuous variations of BPS central charge can be made in a
purely geometric way.  Thus any framework which makes mirror symmetry
manifest must have some way to treat this problem.

In fact, there is such a framework, the homological mirror symmetry
proposal of Kontsevich, which is going to be the primary new example
in our discussion of abstruse mathematics turning into physics.  This
proposal was loosely inspired by Witten's discussion of topological
open string theory \cite{Witten-top}, which identified allowed
boundary conditions in the A and B twisted string theories as
(respectively) isotopy classes of lagrangian submanifolds and
holomorphic bundles.  A mathematical theory of the A objects had been
developed by Fukaya, which had all of the structure required to match
up with B branes as coherent sheaves, but did not.  Rather, Kontsevich
proposed that B branes had to be identified with objects in the
derived category of coherent sheaves.

Again without going into technical details (which are
spelled out to some extent in \cite{DCS} and in more detail in
\cite{AspLaw,Diac}), the basic idea of the derived category is
somewhat analogous to K theory, which is now a generally accepted
element in the discussion of D-branes, in that it allows discussing
arbitrary combinations of branes and antibranes in a precise way.
Where the derived category goes far beyond K theory is that it keeps
track of all massless fermionic open strings between a pair of
D-branes.  This depends on much more than the topological class of the
branes -- for example, a pair of D$0$-branes will come with extra
massless strings only if they are located at the same point.

If we know the massless fermions between a pair of branes, and we
assume $\CN=1$ supersymmetry (which may be spontaneously broken), we
effectively know all of the vector and chiral multiplets in the
world-volume theory which are not lifted by superpotential-induced
mass terms.  In a precise sense, this is all of the holomorphic
information about the branes which does not depend on the stringy
K\"ahler moduli of the CY, and does not depend on which branes are
``branes'' and which are ``antibranes.''  

The appearance of the derived category can also be motivated quite
simply from considerations of topological open string theory: one just
generalizes the definitions to allow $Q_{BRST}$ to contain non-trivial
dependence on the Chan-Paton factors, and then imposes an equivalence
relation which identifies configurations which are related by adding
cancelling brane-antibrane pairs.  The difference with K theory is
that one only considers a brane-antibrane pair to cancel if all open
strings to them entirely cancel out of the $Q$-cohomology, a condition
which requires the brane and antibrane to be identical as holomorphic
objects.  

As is usual in topological string theory, a theory which is
topological on the world-sheet can describe non-topological
information in space-time.  Topologically twisted type \II\ strings on
CY generically contain holomorphic information; the $\CN=2$
prepotential for the closed string, and the $\CN=1$ superpotential for
the open string.  An essential difference between closed and open
strings for the present problem is that the existence of boundary
conditions (D-branes) for the open string problem depends on
non-topological data, the K\"ahler moduli, which drop out of the B
twisted theory.  Thus it is useful to distinguish ``topological''
(one might also say ``holomorphic'') D-branes, whose existence does not
depend on K\"ahler moduli, from ``physical'' D-branes whose existence
does depend on these moduli.

It is important to note that the formalism of the derived category is
general; one can take any ``category'' (obeying certain axioms) and
form the corresponding derived category.  In particular, one can
define categories of configurations of $\CN=1$ supersymmetric gauge
theories, which correspond to branes as discussed above, and then form
the corresponding derived category.  This will lead to a structure
which can describe all bound states both of the original generating
branes $B_i$ and all of their antibranes $\bar B_i$.

There is a lot of mathematical evidence by now that the derived
category is the correct framework for this discussion.  For example,
if one follows a closed loop in K\"ahler moduli space, one obtains a
monodromy on the RR charges of B branes; in fact explicit candidate
transformations on the entire topological brane spectrum have been
proposed \cite{Horja,SeidelThomas}, which act naturally on the derived
category.  Furthermore, the simplest statement of Beilinson's theorem
and the other theorems we mentioned, is that the the derived category
of F-flat configurations of a quiver gauge theory is equivalent to the
derived category of coherent sheaves on the corresponding space.  This
class of objects seems to be large enough to describe all the
mathematical and physical phenomena discovered so far (at least for
BPS branes; non-BPS branes are not so well understood yet) and
combined with the explicit constructions of \cite{DCS,AspLaw,Diac}
we seem to have
adequate physical confirmation of Kontsevich's proposal.

Besides describing all branes in principle, these ideas tell us that
we can find concrete ways to describe all branes in practice -- if we
can find the appropriate quiver gauge theory, its derived category
will be a usable description of all of the branes.  These theories are
known for orbifolds, and quite a lot has already been worked out for
Gepner models which describe compact CY's \cite{DD,GJ,Mayr,Tomas}.
Perhaps the main problem in getting an exact description along these
lines is to work out an exact superpotential for Gepner model boundary
states.

An intriguing conjecture, supported by the known examples, is that
these superpotentials always take a form such that the equations $\p
W/\p \phi_i=0$ are the conditions for an operator $D$ constructed from
the fields $\phi$ to square to zero \cite{toappear}.  A heuristic
argument for this is that this is also true of the equation
(\ref{eq:twozero}).

\section{D-flatness}

Having made what we believe is a correct statement for the set of
``all solutions of F-flatness conditions'' or topological B branes, we
can now go on to discuss the D-flatness conditions.  A direct approach
to this problem would be to find the stringy corrections to the
hermitian Yang-Mills equation, and then either solve the resulting
equation or find the necessary and sufficient conditions for its
solution.  Although there is a better approximation to the ``correct''
equation available (the MMMS equation \cite{MMMS}), this includes only
powerlike $\ap$ and not world-sheet instanton corrections, and it
seems likely that including the latter in a direct approach would be
complicated.  

As we discussed, these equations are already too hard to solve in
the large volume limit, and what is more useful is to know the
necessary
and sufficient conditions under which they will have a solution.
According to the DUY theorems,
a holomorphic bundle on a K\"ahler manifold $M$
will correspond to a solution of hermitian Yang-Mills (which will
be unique) if and only if it is $\mu$-stable.  To define this
condition,
we first define the {\it slope} $\mu(E)$ of a bundle $E$ to be
\begin{equation}
\mu(E) = \frac{1}{\rk E}\int c_1(E)\wedge \omega^{n-1} ,
\end{equation}
where $\rk E$ is the rank of $E$, $\omega$ is the K\"ahler class of
$M$, and $n$ is the complex dimension of $M$.

A bundle $E$ is then $\mu$-stable iff, for all subbundles $E'$, one
has
\begin{equation}
\mu(E') < \mu(E).
\end{equation}
Although the condition is a little complicated to use in practice (one
must be able to work with lists of subbundles, either explicitly or
implicitly), it is both the simplest general mathematical condition
which has been found in work on this subject, and it is physically
meaningful: if one varies the K\"ahler moduli in a way which
causes a degeneration $\mu(E')=\mu(E)$, one can show that one reaches
a line of marginal stability on which the D-brane associated to $E$
will decay into products including $E'$.  Thus the mathematical idea
of ``subbundle'' or ``subobject'' corresponds directly to a physical
idea of ``subbrane,'' which suggests that we should be able to
generalize this condition to any point in stringy K\"ahler moduli
space.

Such a generalization, called $\Pi$-stability, was proposed in
\cite{DFR} and then in a more general form in \cite{DCS}.  The
quantity which plays the role of the slope turns out to be the phase
of the BPS central charge, which is computed from the periods $\Pi$ of
the mirror CY. Important inputs into this proposal were work of Sharpe
on $\mu$-stability and D-flatness \cite{Sharpe}, work of Joyce on A
branes \cite{Joyce} and of Kachru and McGreevy relating this to
D-flatness \cite{KachruMac}, and finally work of King \cite{King}
which (as applied in \cite{DFR}) gives the necessary and sufficient
conditions for D-flatness conditions in $\CN=1$ gauge theory to admit
solutions. Somewhat surprisingly, these last conditions were not
previously known in the physics literature (except for the special
case of zero FI terms).

All of these conditions are interesting, but space prohibits their
discussion here.  In any case, there is a general argument from
world-sheet conformal field theory \cite{DCS} which leads to a general
stability condition which reduces to each of the ones cited above in
the appropriate limit, which we now describe.

Consider the $(2,2)$ superconformal field theory associated to the CY,
with two B-type boundary conditions.  These are discussed in
\cite{OOY}; the important point for us is that they can be regarded as
Dirichlet boundary conditions on the boson representing the $U(1)$ of
the $(2,2)$ algebra, with the position just being the phase of the BPS
central charge.  Two branes which preserve different space-time $\CN=1$'s
preserve the world-sheet $\CN=2$, but spontaneously break space-time
supersymmetry, because the world-sheet boundary conditions eliminate
the zero mode of the spectral flow operator.  Nevertheless,
one can relate space-time bosons to fermions, by using the spectral
flow operator associated to either of the original BPS boundary states.
The results differ by a phase, so this is not a symmetry, but
one can still identify a partner bosonic operator to each massless
fermion (thus each state in the topological open string theory).
Furthermore, one can compute its mass by these considerations; it is still
\begin{equation} \label{eq:mass}
m^2 = \frac{1}{2} (Q-1)
\end{equation}
where $Q$ is the $U(1)$ charge.

This is the standard formula; what has changed is that $Q$ need not
take integer eigenvalues anymore, and bose-fermi masses are not degenerate.
If the two Dirichlet boundary conditions
on the boson are $\grade_1$ and $\grade_2$, the $U(1)$ charge (which is
winding number) will take values 
\begin{equation} \label{eq:qcharge}
Q = n+\grade_2-\grade_1; \qquad n\in\BZ.
\end{equation}

Suppose we now consider a given pair of boundary conditions and open
string, and vary the K\"ahler moduli.  The only effect on this state
(since it is chiral under the world-sheet $\CN=2$) will be to vary
the positions $\grade_1$ and $\grade_2$, and thus the $U(1)$ charge $Q$,
according to (\ref{eq:qcharge}).  

This provides a rule which determines the mass squared of every boson
in a chiral multiplet, everywhere in K\"ahler moduli space, if we know
it at one point.  For B branes, the $U(1)$ charge $Q$ is also the rank
of an associated differential form (for example, $H^1(M,A^*\otimes B)$
corresponds to a charge $1$ chiral primary and massless matter), so this
can be computed from geometry or from the quiver constructions.

This result is the key to understanding bound state formation and
decay: if a boson goes tachyonic, condensing it can form a bound
state, while if it goes massive, a previously existing bound state can
go unstable.  The first conclusion is probably uncontroversial; the
second conclusion can be proven by showing that if at some point in
K\"ahler moduli space two branes $A$ and $B$ formed a bound state $C$
by tachyon condensation; then if at some other point the string
$A\rightarrow B$ becomes massive, some other chiral operator
$B\rightarrow C$ or $C\rightarrow \bar A$ will have its $U(1)$ charge
drop below zero.  However this contradicts the axioms of unitary CFT,
a contradiction which can only be resolved by the decay of the heaviest
of the branes involved.

All of these considerations can be summarized in the following rules.
We need a definition from the formalism of the derived category, the
``distinguished triangle.''  Certain triples of branes are ``distinguished,''
in physical terms because tachyon condensation between a pair of them can
produce the third one as a bound state.  This is denoted by the following
diagram:
\begin{equation} \label{eq:disttri}
\ldots\mapr C[-1]\mapr^\psi A\mapr^\rho B\mapr^\phi C
\mapr A[1] \mapr B[1] \mapr \ldots .
\end{equation}
The arrows denote open strings (called morphisms in the categorical
terminology), while the bracketed notation $A[1]$ indicates an ``image
brane'' as explained in \cite{DCS}; the odd ``images'' are antibranes.

There are various ways to read this diagram: $A$ and $C$ can form the
bound state $B$ by condensing $\psi$; $\bar A$ and $B$ can form $C$ by
condensing $\rho$; and so forth.
In many works (e.g. \cite{HarveyMoore}), it has been noted that bound
state formation can be described by exact sequences; for example $A +
C$ binding to make $B$ is
\begin{equation}
0 \mapr A\mapr B\mapr C \mapr 0 .
\end{equation}
Every such exact sequence leads to the corresponding distinguished
triangle in the derived category, describing various related
brane-antibrane processes.  Furthermore, every pair of objects
(branes) and every morphism between them can be completed to a
triangle, so there are many more possible bound states in this
framework.

To find lines of marginal stability or predict bound state formation,
one works with these triangles, and keeps track of the $U(1)$ charge
or ``grading'' of each of the three open strings involved.  By the
definitions, these will always sum to $1$.  One then enforces the
basic rule that no chiral field can have negative $U(1)$ charge.
Thus, if all three objects in the triangle actually exist as physical
branes (are stable) at a given point in K\"ahler moduli space, all
three strings must have $U(1)$ charges between $0$ and $1$.  We refer
to this as a ``stable triangle.''  Conversely, if two stable objects
are related by an open string with $Q>1$, the third object in the
triangle must not be stable.

Suppose we have all this information and we then follow some path in
K\"ahler moduli space.  The $U(1)$ charges of the open strings will
vary, and when they reach $0$ or $1$, decays are possible.  More
specifically, if some $U(1)$ charge for a stable triangle becomes $0$,
one can show that the others must be $0$ and $1$ (this follows from the
relation $Z_A + Z_C = Z_B$ between the central charges).  As we cross
this line, the brane between the $0$'s will decay (it will always be
the heaviest of the three).  On the other hand, suppose we start with
two stable objects with a massive open string $Q>1$ which becomes massless.
As we cross this line, the third object (their bound state) will go stable.

There are many consistency conditions which these rules must satisfy.
For example, in the second process, one requires that the new bound state
is not destabilized by some other object.  Furthermore there are relations
required to make this rule unambiguous; for example it cannot be that
an object both enters into bound state formation and decays on the same
line.  Some but not all of these conditions have been proven from the
formalism at present.  Furthermore, many nontrivial examples seem
to make sense, and as we mentioned one can rederive the known stability
conditions in appropriate limits, so this rule appears to be a good
candidate for the necessary and sufficient condition replacing the
D-flatness conditions on a general stringy Calabi-Yau.

\section{An example}

The simplest example of all of this is the formation of a D$p-2$-brane
by tachyon condensation between a D$p$-brane and an anti-D$p$-brane
carrying flux, as discussed by many authors (in the CY context, by
\cite{OPW}).  This is described by the exact sequence
\begin{equation}
0 \mapr \CO(-1) \mapr^0 \CO \mapr^{1/2} \CO_\Sigma \mapr 0 .
\end{equation}
Here $\CO$ is the D$p$-brane with trivial gauge bundle wrapping the entire
CY or some compact cycle in it.  $\CO(-1)$ is a brane carrying $-1$ unit
of flux, and $\CO_\Sigma$ is the resulting D$p-2$-brane.

We have indicated the $U(1)$ charges of the maps involved, computed in
the large volume limit, by the numbers above the arrows.  The
$\mapr^0$ indicates a standard brane-antibrane tachyon with
$m^2=-1/2$, while the $\mapr^{1/2}$ indicates a D$p$-D$(p-2)$ tachyon
with $m^2=-1/4$ (by the usual large volume rules); the charges are
then deduced using (\ref{eq:mass}).  If we complete this to a triangle
as in (\ref{eq:disttri}), the third map will also have charge $1/2$.

Now, to see what happens as we decrease the volume of the CY, we need
to know how the BPS central charges of these branes vary.  This can be
computed from mirror symmetry, but in examples studied so far it turns
out that the qualitative behavior we are interested in is already
predicted by the results with world-sheet instanton corrections left out.
In this case, we have (for a cycle of dimension $n$)
\begin{eqnarray}
Z(\CO) &= \frac{1}{n!}(B-iV)^n; \\
Z(\CO(-1)) &= \frac{1}{n!}(-1+B-iV)^n; \\
Z(\CO_\Sigma) &= Z(\CO) - Z(\CO(-1)).
\end{eqnarray}
As we decrease $V$, $\grade(\CO)$ will increase, and $\grade(\CO(-1))$
will decrease, so the charge $0$ will increase, while the $1/2$ charges
can be checked to decrease.  Eventually the first charge reaches $1$, and
the brane $\CO_\Sigma$ will decay into these two constituents (assuming
it didn't decay into something else first).\footnote{The same argument
predicts that a higher degree D$p-2$-brane will also decay \cite{DW}, 
resolving a paradox encountered in \cite{MSW}.}
Thus this brane does not
exist in the small volume region.  For reasons we will not get into 
here, it is natural to identify this marginal
stability line with a ``phase boundary.''

This conclusion would presumably follow from any correct treatment of
the bound state.  However, an amusing and nontrivial prediction of the
present formalism is that if one continues to smaller volume, another
bound state forms.  This is because the map $\CO(-1)\mapr \CO$ will
have a Serre dual, which on a Calabi-Yau will be an element of
$H^n(\CO,\CO(-1))$.  (This is the dual under the natural pairing
$(\alpha,\beta)=\int \bar\Omega^{(n)} \wedge \Tr \alpha \wedge \beta$.)
This fits into another triangle
\begin{equation}
\ldots \mapr \CO \mapr^n \CO(-1) \mapr X \mapr \CO[1] \mapr\ldots 
\end{equation}
where the object $X$ is not a coherent sheaf, and is not stable at
large volume.

Now, the same considerations which made the $U(1)$ charge of $\mapr^0$
increase with decreasing volume, will make the $U(1)$ charge of $\mapr^n$
decrease.  Eventually it will cross $1$ (at some smaller volume
than the previous decay) and $X$ will become a new stable brane.

In the Gepner model of the quintic, one can check that this happens
before one reaches the Gepner point.  In fact, $X$ turns out to be one
of the states constructed by Recknagel and Schomerus, the
``mysterious'' state discussed in \cite{Doug,Denef} whose central
charge would have vanished on a path leading back to large volume.
Thus we have an example of a brane which does not correspond to a
coherent sheaf yet which has been proven to exist in the stringy
regime, which serves as a nontrivial confirmation of these ideas.


\end{document}